\documentclass[10pt, twocolumn]{article}	

\usepackage[utf8]{inputenc}
\usepackage{amsfonts, amsthm, amssymb, mathtools}
\usepackage[caption=false]{subfig}
\usepackage{tikz}
\usepackage{graphicx} 
\usepackage{authblk}
\definecolor{myurlcolor}{rgb}{0,0,0.7}
\definecolor{myrefcolor}{rgb}{0.8,0,0}

\usepackage{hyperref}

\definecolor{ppblue}{RGB}{46,117,182}
\definecolor{ppred}{RGB}{197, 90, 17}

\DeclareMathOperator*{\argmax}{arg\,max}

\def\>{\rangle}
\def\<{\langle}

\begin{document}
\title{Uncertainty of Feed Forward Neural Networks Recognizing Quantum Contextuality}



\author[1]{{Jan Wasilewski} 
\footnote{Correspondence: wasilewskij823@gmail.com}
\footnote{Code available on \href{https://github.com/JohnnyWasilewski}{https://github.com/JohnnyWasilewski}}}
\author[2,3]{Tomasz Paterek }
\author[4,5,6]{Karol Horodecki}
\affil[1]{Institute of Informatics, Faculty of Mathematics, Physics and Informatics, University of Gdańsk, Wita Stwosza 57, 80-308 Gdańsk, Poland}
\affil[2]{Department of Physics, School of Mathematics and Physics, Xiamen University Malaysia, 43900 Sepang, Malaysia}
\affil[3]{Institute of Theoretical Physics and Astrophysics, Faculty of Mathematics, Physics and Informatics, University of Gdańsk, Wita Stwosza 56, 80-308 Gdańsk, Poland}
\affil[4]{Institute of Informatics, National Quantum Information Centre, Faculty of Mathematics, Physics and Informatics, University of Gdańsk, Wita Stwosza 57, 80-308 Gdańsk, Poland}
\affil[5]{International Centre for Theory of Quantum Technologies, University of Gdańsk, Wita Stwosza 63, 80-308 Gdańsk, Poland}
\affil[6]{School of Electrical and Computer Engineering, Cornell University, Ithaca, New York 14850, USA}

\setcounter{Maxaffil}{0}
\renewcommand\Affilfont{\itshape\small}
\date{\today}
\maketitle
\textbf{ 
The usual figure of merit characterizing the performance of neural networks applied to problems in the quantum domain is their accuracy, being the probability of a correct answer on a previously unseen input.
Here we append this parameter with the uncertainty of the prediction, characterizing the degree of confidence in the answer.
A powerful technique for estimating both the accuracy and the uncertainty is provided by Bayesian neural networks (BNNs).
We first give simple illustrative examples of advantages brought forward by BNNs, out of which we wish to highlight their ability of reliable uncertainty estimation even after training with biased data sets.
Then we apply BNNs to the problem of recognition of quantum contextuality which shows that the uncertainty itself is an independent parameter identifying the chance of misclassification of contextuality. 
}


		
Machine learning is one of the key techniques of data processing in modern information-based technology. Everyday applications of feed-forward neural networks (NNs) range from speech \cite{speech} and image \cite{computervision} recognition to computer-aided medicine \cite{medicine}, and autonomous vehicles \cite{av}.
Machine learning also reached quantum information theory, among others, by harnessing NNs to solve some intrinsically quantum problems \cite{Bharti_2020}. A number of these problems can be cast as resourceful object recognition \cite{Bharti_2020,Chitambar_2019}. The task is to train the network~\cite{dl} by exposing it to objects which are ``free'' i.e. resource-less as well as ``non-free'' i.e. resourceful, and later verify its predictions in telling if a given previously unseen object contains a resource such as non-locality \cite{Bell-nonlocality,Canabarro_2019,Kriv_chy_2020}, entanglement \cite{Horodecki_2009,Entanglement_DNN_recognition} or steering \cite{Uola_2020,Zhang_2021}. 
The percentage of cases in which the answer is correct is called the accuracy of the network.
Other contexts in which machine learning has been used to aid quantum information science include but are not limited to  finding quantum super-additive channels \cite{Banik}, finding quantum states with specific entanglement structure (e.g., bound entangled states) \cite{Hiesmayr_2021}, or
aid in tomography \cite{tomography,Ghosh2021}.

Another key notion in theoretical science, including quantum information theory, is that of uncertainty, here quantifying confidence in the network's answer.
Since a trained NN is an ``oracle'', due to its complicated structure, it is desirable to know if its output is trustworthy.
For example, if the network is asked about a truth value of a certain proposition, a confident answer could be seen as a theorem, whereas a less confident one as a conjecture.
If the problem at hand is of high risk, such as regarding medical conditions or transport safety, the uncertainty estimations are essential because we would like a human controller to take over when the network is unsure of its answer.
Note that this holds even for highly accurate networks.
While uncertainty estimations in the classical domain are a topic of intensive research \cite{Un_review, uncertaitny}, it is still in its infancy for quantum-aided networks~\cite{QBNN} and has not yet been tested in the case of the NNs harnessed for quantum information-theoretic problems.

Here we initiate uncertainty estimation for NNs applied to quantum problems.
We first provide illustrative examples of different methods for uncertainty estimations aimed at a reader from the physics community.
We compare the standard way of computing the uncertainty of NNs with estimations from Bayesian neural networks (BNNs), which are probabilistic models of neuromorphic data processing \cite{MacKay1, MacKay2}.
In particular, we demonstrate that BNNs have superior uncertainty estimations when training data is limited and biased, a fact of interest also to the neural network community.
We then apply these methods to the problem of recognizing quantum contextuality~\cite{Kochen-Specker, Joshi_bound, Peres_Incompatible, Mermin_quantum_probability,Contextuality_review}.
In a nutshell, contextuality is a phenomenon of nonexistence of a hidden variable model where measurement results are well-defined before the measurement and are independent of other measurements conducted simultaneously (the context). 
The roots of contextuality are in the inherently quantum-mechanical interdependencies expressed by (non)commutation relations of the corresponding observables.
It has been found recently that contextuality empowers solutions to many problems, including classical communication problems \cite{CLMW,Leditzky2020,Saha2019}, quantum device-independent cryptography \cite{Context_crypto}, and quantum computing \cite{Howard_etal, BermejoVega2017,Bravyi2018}.
Therefore the classification of phenomena into contextual vs. noncontextual is an important task. 
This task can be seen as an instance of resourceful object recognition, which has not yet been explored by machine learning techniques. See Ref.~\cite{ai_forphysics} for a survey of other foundational applications of neural networks.
We focus here on a simple scenario exhibiting contextuality, which has already been independently characterized, in order to demonstrate the uncertainty estimations in a clear and systematic way \cite{ncycle}.


\section{Uncertainty in neural networks}

Neural networks are parameterized models consisting of sequences of layers of neurons. Suppose the first layer consists of $m$ neurons. 
The action of these neurons on an $n$-dimensional input data vector is a multiplication by a $n \times m$ matrix $w_1$ followed by the addition of  $m$-dimensional bias vector $b_1$.
Such obtained output is operated on by the nonlinear element-wise transformation called the activation function. The resulting vector is then multiplied by the weight matrix of the second layer $w_2$ and the procedure is repeated until the last layer is reached. The network parameters are usually collectively referred to as the weights $w = \{w_l, b_l\}_l$. 

For classification tasks, where the algorithm must assign a class to the input, the number of neurons of the last layer is equal to the total number of classes. 
The softmax function is then applied in order to normalize the output vector in the following way:
\begin{equation}
\text{softmax}_j (z) = \frac{e^{z_j}}{\sum_i e^{z_i}},
\end{equation}
where $z$ is the output vector with components $z = (z_1, \dots, z_d)$, i.e. the total number of classes is $d$.
A function $\text{softmax}(z)$, without an index, returns $d$-dimensional vector with entries given by $\text{softmax}_j (z)$.
The weights and biases of each layer are established by minimizing the loss function during the process called training. In the case of the classification, this loss function is usually chosen as the cross-entropy:
\begin{equation}
\label{EQ_CROSSENT}
L(w|D) = -\sum_{i=1}^N \log (\mathrm{NN}(x_i, w)_{y_i}),
\end{equation}
where $\mathrm{NN}(x_i, w)_{y_i}$ is the
output of neural network with weights $w$ applied on input vector $x_i$ which belongs to the class $y_i$.
Note that in the supervised learning, each input is already ascribed to its class, which is represented by a dataset $D=\{x_i, y_i\}_{i=1}^N$, where $N$ is the size of the training set. 
The minimization of the loss function is usually done by the algorithm called gradient descent \cite{gradient}, which starts from a random $w$ and iteratively updates it in the direction of the lowest gradient, calculated by the algorithm called backpropagation \cite{backpropagation}. Despite being very powerful and useful for practical applications, this way of training the neural network has its limitations as described below.

\subsection{Standard neural networks}

Although present-day neural networks are already extremely accurate, they will never reach perfect accuracy for nondeterministic noisy tasks. The limited amount of training data usually results in the necessity to capture the patterns from incomplete information. The other source of inaccuracy is the fact that the numerical methods minimizing the loss function can converge to a local minimum.
The incredible success of NN models can be theoretically motivated by the universal approximation theorem, which states that any continuous relationship between $x$ and $y$ can be approximated by some neural network model~\cite{UAP}. 
Note, however, that this is an existence statement and does not hint at any concrete NN architecture for a given concrete problem.
In practice, different problems are solved better by different architectures, leading to yet another source of inaccuracy.
Recognizing in real time the moment when a given network may result in an incorrect prediction is the task of uncertainty estimation, which is our focus from now on.

A common method of uncertainty estimation in standard NN proceeds as follows.
The output of the network is a normalized vector with a dimension equal to the number of classes. The final prediction is done by taking the class with the highest score. 
The natural way of thinking about the output vector is that it represents the probabilities of the input belonging to every class. Therefore the intuition is that the prediction $[0.1, 0.9]$ in the binary classification problem
is less uncertain than the prediction $[0.4, 0.6]$ though it is the second class that is predicted in both cases. The uncertainty can be calculated as the entropy of the components of the output vector
\begin{equation}
E(p) = - \sum_{c=1}^d p_c \log p_c.
\end{equation}
The experiments below and those in the literature \cite{overconfidence} report that NNs with such uncertainty estimation are overconfident in their predictions. 
The output vectors admit low entropy also for 
the inputs far from the points in the training dataset, i.e., out-of-distribution samples, and even for samples improper for the specific task (for example, in the case of image classification ``cat versus dog'', when the input picture contains the bird). 
The overconfidence is clearly seen for input spaces not entirely covered by the training dataset. This undesired feature makes  uncertainty calculated as the entropy of the output of little use.

There exist many alternative ways to
reduce the overconfidence just described, see \cite{Un_review} for a comprehensive review, where  methods to recognize the moment when the output becomes uncertain are also presented.
Among them, Bayesian neural networks is the technique of our choice here.

\subsection{Bayesian neural networks}

Bayesian neural networks (BNNs) were proposed by MacKay in Refs.~\cite{MacKay1,MacKay2}.
They enable inference of the uncertainty of a given model of a deep neural network via Bayesian statistical methods. The basic idea behind the uncertainty of neural networks in the Bayesian approach is that it allows for distribution over the set of networks with a given architecture. We infer if the network is confident by comparing outputs of the sampled network realizations on the given input.

In order to define BNN model, we shall now give a more technical, probabilistic description of neural networks together with their use on concrete illustrative examples.
Consider the supervised learning problem of classification of inputs into particular classes.
Let us label the training data $\{x_i, y_i\}$, where $x_i$ is the input, and $y_i$ is the output. Index $i$ runs over the training set. A fixed size neural network with weights $w$ is a  function NN$(\cdot, w):\mathcal{X} \rightarrow \mathcal{Y}$ which approximates true relation between $x$ and $y$ by using weights $w$ in different layers. We assume probabilistic model of $y$ given $x$:
\begin{equation}
p(y_i|x_i, w) = \text{softmax}(\mathrm{NN}(x_i, w)).
\end{equation}
Since the samples are statistically independent, the probability distribution of the dataset $D$ is $p(D|w) = \prod_{i=1}^N p(y_i|x_i, w)$. Frequentist estimation of the parameters $w$ can be done by maximizing the log-likelihood
\begin{equation}
w_0 = \argmax_w \log p(D|w),
\end{equation}
which is equivalent to minimizing the cross-entropy loss function given in Eq.~(\ref{EQ_CROSSENT}), see, e.g., \cite{murphy}.

Bayesian statistics provides a different way of estimating the parameters of the model, having the advantage of providing the whole distribution over the weights rather than just point estimates, as in the case of frequentist estimation. We first assume \emph{a priori} distribution over the weights $p(w)$. 
Practically it is often chosen as a 
multidimensional normal distribution (centered at zero) with a fixed diagonal variance \cite{priors}. Bayes theorem then provides an update of our belief about the weights:
\begin{equation}
p(w|D) = \frac{p(D|w)p(w)}{p(D)}.
\end{equation}
This distribution is called \emph{a posteriori} distribution, and in general, it is intractable due to the unknown distribution $p(D)$. 
Yet, one can approximate $p(w|D)$ via different methods.
We shall use a variant of the so-called Monte Carlo Markov Chain (MCMC), which samples from $p(w|D)$ and produces iteratively the sequence of $w$'s, which are theoretically guaranteed to converge to $p(w|D)$.
The trick used is that in the $i$th iteration, in order to generate the next $w_{i+1}$, it is enough to calculate the ratio $\alpha=p(w_i|D)/p(w'_{i+1}|D)$, where $w'_{i+1}$ is the candidate next set of network weights and in this way the dependence on $p(D)$ cancels out.
Based on the $\alpha$ value the candidate is either rejected and $w_{i+1} = w_i$ or it is accepted $w_{i+1}=w'_{i+1}$. There are many ways of calculating the candidate $w_{i+1}'$, the easiest being $w'_{i+1}=w_i + \epsilon$, where $\epsilon$ is from a normal distribution centered at zero.
In practice, the dimension of $w$ is very high, and random choice of the direction which carries more probable parameters is very unlikely, resulting in the rejection of the candidate most of the time. We, therefore, use a more sophisticated way of proposing the candidates, motivated by the physical arguments from Hamiltonian dynamics.
The resulting method is called Hamiltonian Monte Carlo (HMC) \cite{mcmc_nn, HMC}.
Having $m$ significant weights $\{w_i\}_{i=1}^m$~\footnote{First couple of samples are usually rejected due to the fact that the initial $w$ are chosen at random}, sampled from the posterior distribution, the predictive distribution is approximated as:
\begin{equation}
p(y|x) = \int p(y|x, w) \mathrm{d} p(w|D) \approx \frac{1}{m}\sum_{i=1}^m p(y|x, w_i)
\end{equation}
where $p(y|x)$ is the probability
that the trained BNN represented by the sequence of weights $\{w_i\}_{i=1}^{m}$, on input $x$ will output $y$.
As we will see, such predictions possess more informative uncertainties.


\subsection{Illustrative example}

\begin{figure*}[!t]
\begin{flushleft}
\textbf{Panel a}
\end{flushleft}
    \includegraphics[width=\textwidth, height=6cm]
    {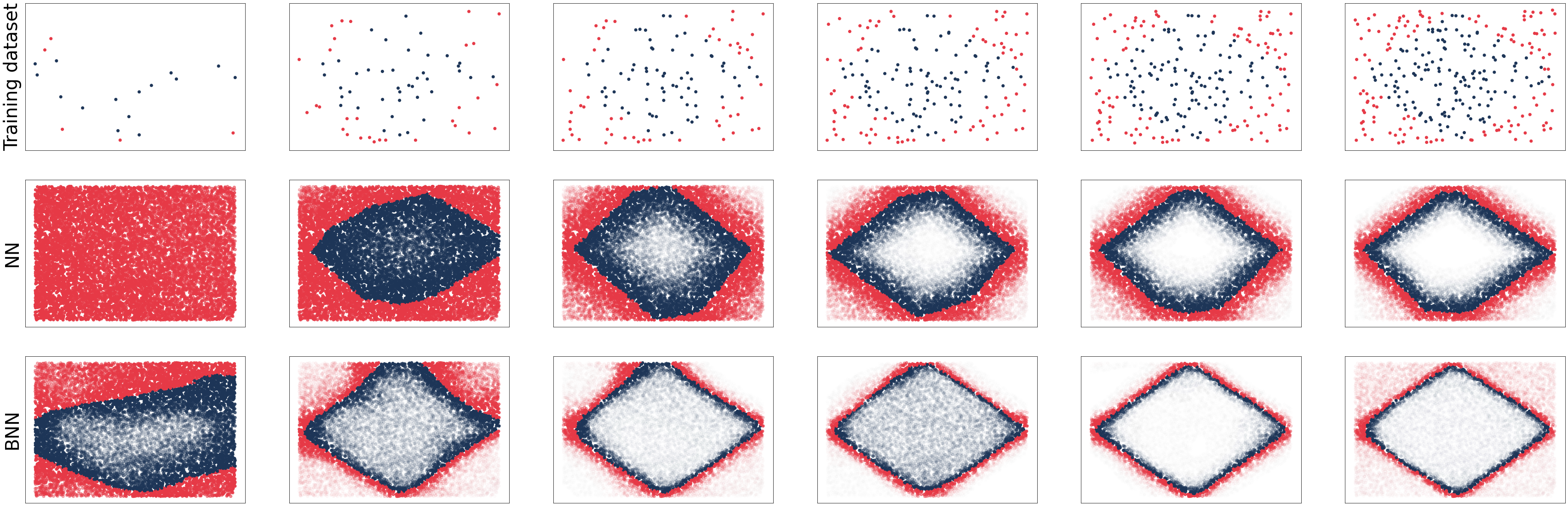}
\begin{flushleft}    
 \textbf{Panel b}   
\end{flushleft}
    
    \includegraphics[width=\textwidth, height=6cm]{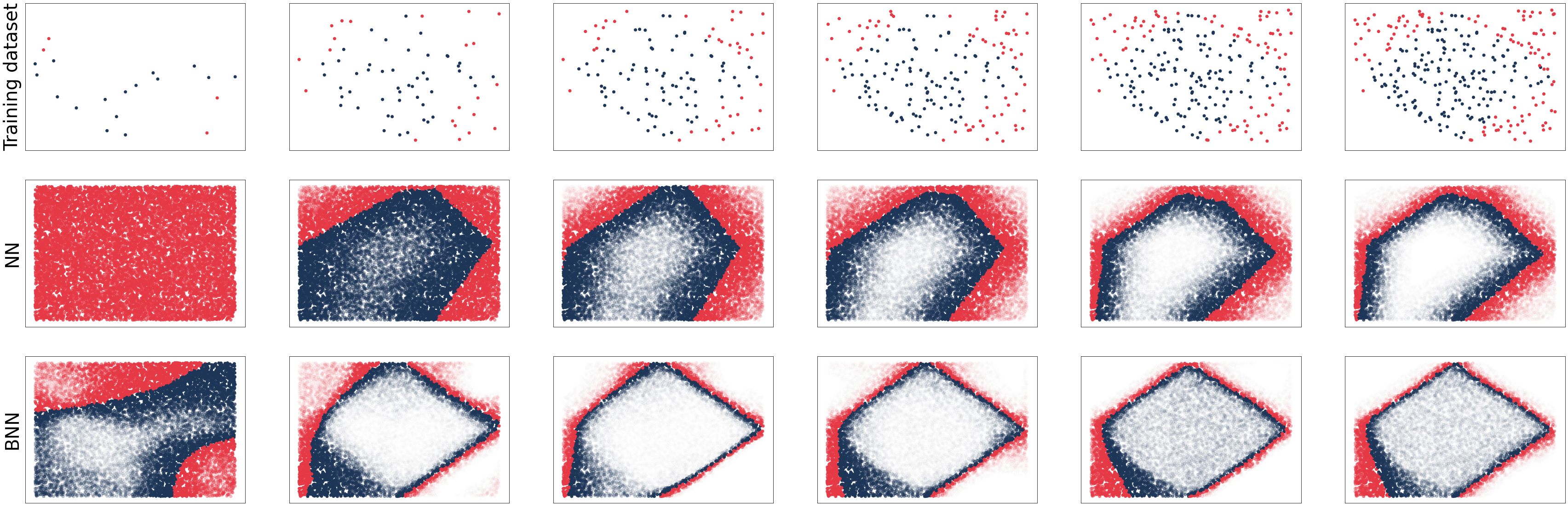}
    \caption{Uncertainty of standard and Bayesian neural networks with three layers containing [8,4,2] neurons. (a) Both networks are trained on a uniformly random distribution. The first row shows the training data, the second row gives predictions of the standard neural network, and the third row the predictions of the Bayesian neural network. The number of data points in the training set grows along the horizontal axis. Blue points indicate the prediction that the point belongs to the rhombus, while red dots indicate the prediction of the outside region. The level of transparency describes the uncertainty --- more transparent predictions are more uncertain.
    (b) The networks are trained with a biased set as showed in the first row. The sampling density of the lower left corner is 50 times smaller than all the other regions. Other notation as in panel (a). }
    \label{FIG_ILLU}
\end{figure*}

Many concrete problems boil down  to classifying whether an object belongs to a convex body or not. As an illustration, we choose a two-dimensional classification scenario to compare BNN and NN behaviour in a simple classification task. 
We sample training data from the square $\mathbb{S} = [-1,1]^2$, which is divided into two regions (two classes). The first region contains points in the rhombus with vertices $(\pm 1,0)$ and $(0,\pm 1)$, whereas the second class contains all other points in $\mathbb{S}$. The task is to recognize the points inside the rhombus.
In Fig.~\ref{FIG_ILLU}, we show that both types of neural networks learn in a similar way, but BNNs are more conservative about their uncertainty estimations.
A vivid case is presented in panel (b), where the networks are trained with biased data.
Clearly, the standard NN are overconfident in their answers and stick to low uncertainty even when they are wrong (see the predictions for the lower left corner with less dense sampling in training). This motivates us to use BNNs in real-world experiments, where the training data is seldom sampled uniformly at random.


\section{Recognising contextuality}
We propose the usage of BNNs to the problem of recognizing quantum contextuality. 
Up to our knowledge, this problemis tackled with the help of NN for the first time.
The network we use is classical, in particular, it is implemented on a classical computer.
It is supplied with so-called behaviour, which is a conditional probability distribution (of measurement results given suitable settings),
and its goal is to predict whether the input behaviour is contextual or not.
In a sense, the network is a witness of contextuality and could be compared to a non-contextual inequality~\cite{Contextuality_review}. Our network is trained in a Bayesian way, and in addition to classifying the input, it also returns the uncertainty of this classification. 
It turns out that such obtained uncertainty acts as we expect: when certain yields correct prediction most of the time, and when uncertain, makes more or less a random guess.


\subsection{Contextuality prediction}

Contextuality is a property of conditional probability distribution (behaviour) of measurement results given measurement settings.
The behaviour is contextual if it cannot be obtained as marginals of a joint probability distribution of outcomes for all possible measurement settings, see e.g.~\cite{Markiewicz2014}.
Contextuality has also been characterized in terms of resource theory in Refs.
\cite{Acn2015,Karol_Axiomatic}.

As a set of behaviors to be classified as contextual or non-contextual by the network, we choose the conditional probabilities related to so-called $n$-cycle~\cite{ncycle}. The case of $n=4$ corresponds to the famous CHSH scenario~\cite{CHSH}, but we focus on $n=5$ as the simplest different case from Bell non-locality (note that Bell inequalities have already been studied with neural network methods~\cite{nn_bell, nn_bell2}).
For $n=5$ the non-contextual behaviors are characterized by the KCBS inequalities discovered in \cite{KCBS}.
These inequalities involve a set of five dichotomic observables $B_i \in \{-1, +1\}$ arranged in the following cyclic way $\{(B_0,B_1), (B_1,B_2), \ldots , (B_4, B_0 )\}$.
Each pair consists of two commuting, i.e., co-measurable, observables, and such pairs are called contexts.
Note that there are only four possible outcomes for each pair $\{(+,+), (+,-), (-,+), (-,-)\}$. The behaviour is represented as a conditional probability table:
\begin{center}
\begin{tabular}{ |c|c|c|c|c|c| } 
\hline
 & $B_0B_1$  & $B_1B_2$  & $B_2B_3$  & $B_3B_4$  & $B_4B_0$ \\
 \hline
 $++$ & $p_{00}$ & $p_{01}$ & $p_{02}$ & $p_{03}$ & $p_{04}$ \\ 
 \hline
$+-$ & $p_{10}$ & $p_{11}$ & $p_{12}$ & $p_{13}$ & $p_{14}$ \\ 
 \hline
$-+$ & $p_{20}$ & $p_{21}$ & $p_{22}$ & $p_{23}$ & $p_{24}$ \\ 
 \hline
$--$ & $p_{30}$ & $p_{31}$ & $p_{32}$ & $p_{33}$ & $p_{34}$ \\
 \hline
\end{tabular}
  \end{center}
Here $p_{ij}$ gives the probability that outcomes measured in the $j$th context ($j=0,\dots,4$) are given in the first column of the $i$th row.
For example, $p_{12}$ is the probability that the results are $+-$ in the measurement of $B_2 B_3$.
There are 10 independent parameters characterizing the behaviour which can be expressed as expectations of individual observables $\langle B_i \rangle$ and correlators $\< B_jB_{j+1} \> = p(+,+) + p(-,-) - p(+,-) - p(-,+)$ (here and below addition $j+1$ is taken modulo $5$) \cite{ncycle}. 
The behaviour is therefore written as $\mathcal{B} = \{\<B_0\>,\ldots,\<B_4\>, \<B_0B_1\>, \ldots, \<B_4B_0\>\}$. 
For the behaviour to be meaningful, all the probabilities in the table above must be non-negative, which means~\cite{ncycle}:
\begin{equation}
      \begin{array}{c}
4p_{1j} = 1 + \< B_j \> + \< B_{j+1} \> + \< B_jB_{j+1} \> \geq 0 \\
4p_{2j} = 1 + \< B_j \> - \< B_{j+1} \> - \< B_jB_{j+1} \> \geq 0 \\
4p_{3j} = 1 - \< B_j \> + \< B_{j+1} \> - \< B_jB_{j+1} \>\geq 0 \\
4p_{4j} = 1 - \< B_j \> - \< B_{j+1} \> + \< B_jB_{j+1} \>\geq 0 \\
  \end{array}
  \label{nondisturbence}
\end{equation}
\begin{figure*}[!t]
    \centering
    \includegraphics[width=\textwidth, height=5cm]{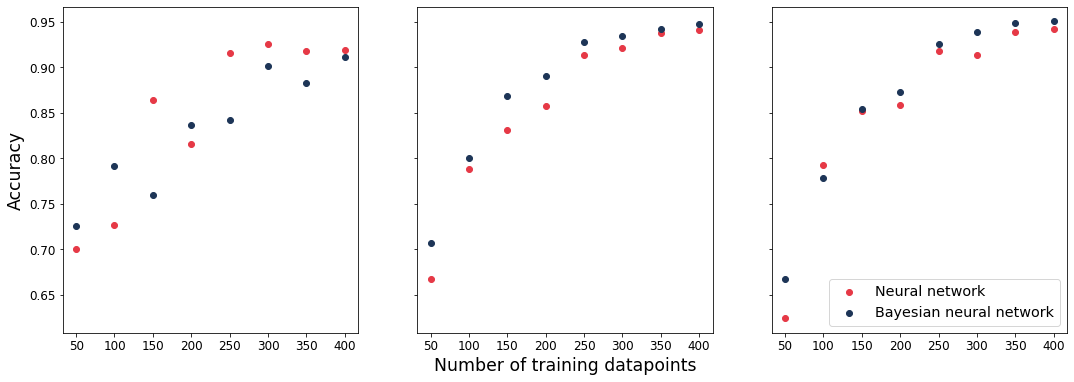}
    \caption{Accuracy of NN (red) and BNN (blue). Each panel shows the accuracy as a function of the size of the training dataset. Different panels give results for different architectures:
 a) a five-layer network with [128,64,32,16,2] neurons, b) a four-layer network with [64,32,16,2] neurons, and c) a three-layer network with [32,16,2] neurons. The results show that both models learn with the same accuracy.}
    \label{fig:acc}
\end{figure*}
The behaviours $\mathcal{B}$ satisfying (\ref{nondisturbence}) automatically satisfy so-called consistency condition (an analog of a non-signaling condition in Bell scenario)~\cite{contex}. These behaviours form a  polytope that contains two separate sets: a non-contextual polytope and a contextual set, which is the difference between consistent and non-contextual polytope.
The behaviour belongs to the non-contextual polytope if it satisfies the inequalities~\cite{ncycle}:
\begin{equation}
\sum_{j=0}^4 \gamma_j \<B_jB_{j+1}\> \leq 3,
\label{context}
\end{equation}
where $\gamma_j \in \{-1, 1\}$ such that the number of $\gamma_j =-1$ is odd.
\begin{figure}
    \includegraphics[width=0.9\linewidth, height=0.9\linewidth]{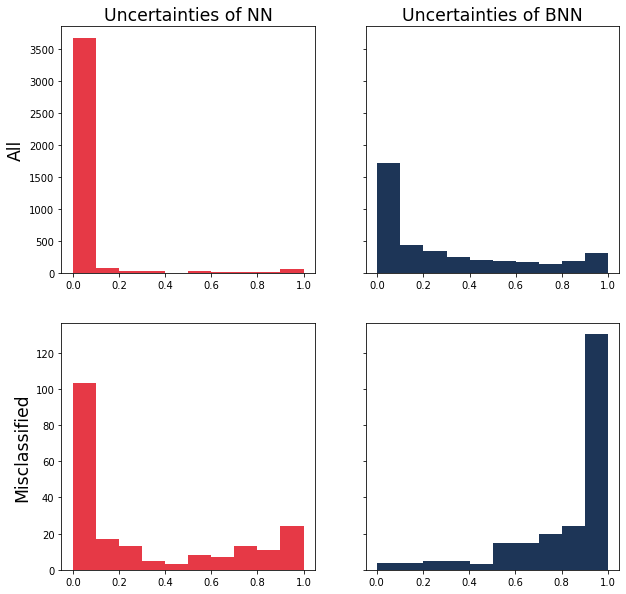}
    \caption{Uncertainty of NN (red, left 
    panels) and BNN (blue, right panels). 
    Histograms of uncertainties of all 
    predictions (top row) and wrong 
    predictions (bottom row) of NN (left column) and 
    BNN (right column). 
    Both models consisted of layers with 
    [64,32,8,2] neurons and were trained on 
    500 training samples and tested with 
    4000 behaviours. Note the overconfidence 
    of NN, which are almost always certain, 
    even when they are wrong.}
    \label{histogram}
\end{figure}

In order to compare the performance of BNN and NN on our problem, we train both models on simulated datasets. 
The training and testing datasets are obtained by uniformly sampling 10-dimensional vectors with entries in the range $[0,1]$ and then rejecting the vectors that do not satisfy (\ref{nondisturbence}). 
In the training set, the labels were assigned by checking non-contextuality conditions (\ref{context}). 
We investigate the accuracy and uncertainty as a function of the growing size of  the training set.
Fig.~\ref{fig:acc} presents the results showing that both types of networks learn with essentially the same accuracy given the training dataset size.
As expected, both models achieve better accuracy on bigger training datasets.

Fig.~\ref{histogram} shows the histograms of uncertainty estimations by both types of networks. The main implication of the performed experiments
is a high correlation between certain and accurate predictions of BNN and the absence of this correlation in NN.
This conclusion is independent of the architecture, i.e., the number of layers and neurons in each layer.
Standard neural networks misclassified samples on which there were very highly confident. In contradistinction, BNN is more ``careful'' with its predictions. Therefore, high uncertainty of BNN gives a hint that the prediction about a given behavior might be wrong. This shows that uncertainties obtained from BNNs are informative and help in the decision process.

\subsection{Aleatoric and epistemic uncertainty}

Bayesian neural networks, in addition to having more informative uncertainty predictions, give the possibility to identify its two main components: aleatoric and epistemic uncertainty \cite{ale_epi}. The former captures the intrinsic uncertainty in the data due to overlapping classes, as illustrated by the task of predicting gender based on human height. 
Since the decision must be based only on the height, the best accuracy would be provided by a decision rule predicting male gender if height is larger than, say, $170$ cm and female otherwise. However, there are females taller than $170$ cm who will be misclassified. Therefore, there is an uncertainty associated with the nature of the problem itself. This type of uncertainty is irreducible, i.e. it can not be decreased by any amount of additional training data. Formally, the aleatoric uncertainty is defined as the integral over a posteriori distribution from the entropy of likelihood
\begin{equation}
A(y,x) = \int H(p(y|x, w))dp(w|D).
\end{equation}
The remaining uncertainty is called epistemic and comes, among others, from the choice of the specific architecture of the neural network and finite set of training samples:
\begin{equation}
 E(y,x) = H(p(y|x)) - A(y,x).
\end{equation}
The epistemic uncertainty can be reduced by collecting more data samples. 



\begin{figure}
    \centering
    \includegraphics[width=0.995\linewidth]{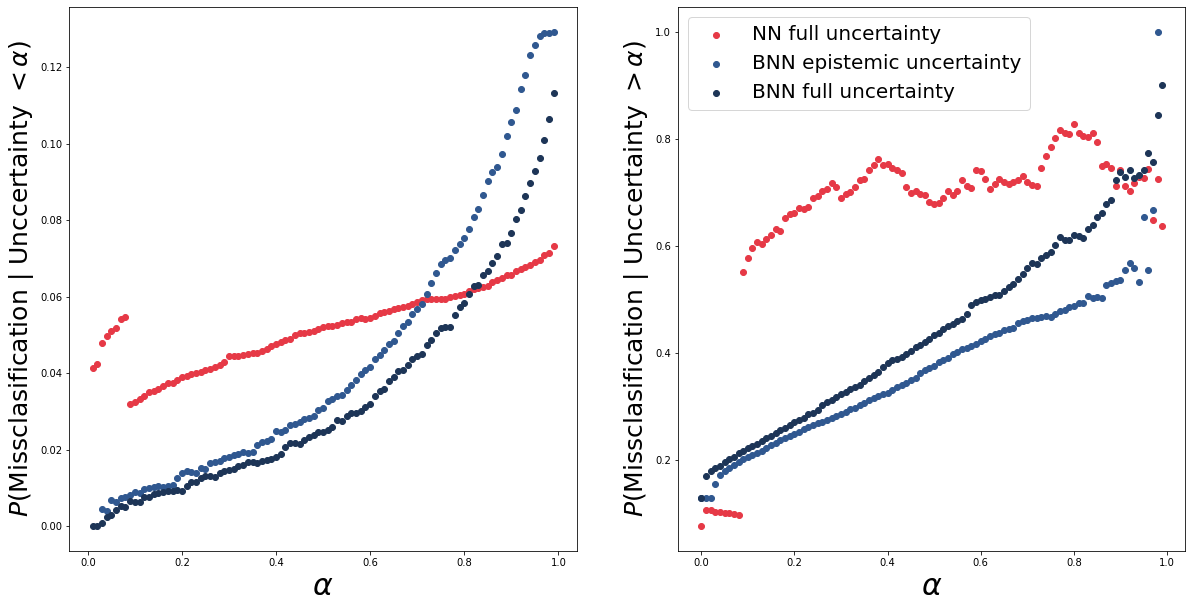}
    \caption{Probabilities of misclassification conditioned on the value of uncertainty. NN and BNN models were trained on contextuality prediction tasks with 500 training data with the architecture [64,32,8,2].}
    \label{fig:my_label}
\end{figure}
We now examine how much information about possible errors in predictions is provided by the estimated uncertainties and what is the role of the discussed uncertainty components. To this aim we introduce conditional probability distributions that characterise misclassification $M$ given uncertainty $U$ obtained from the predictions. 
We approximate these probabilities by dividing the number of misclassified and uncertain samples by the number of suitable uncertain samples and distingish two cases.
The case of ``high'' uncertainty:
\begin{equation}
P(M |U > \alpha) \approx \frac{\#(M \text{ AND } U > \alpha)}{\# (U > \alpha)},
\end{equation}
 and the case of ``low'' uncertainty:
 \begin{equation}
 P(M | U < \alpha) \approx \frac{\#( M \text{ AND } U < \alpha)}{\# (U < \alpha)}.
\end{equation}
The results for NN, BNN and only epistemic uncertainty of BNN are presented in Fig.~\ref{fig:my_label}. 
The misclassification probabilities of NN lie within about $40\%$ of their relevant range (from 0 to $1/2$) and their values do not uniquely correlate with the uncertainty. On the other hand, the misclassification probability is more or less a quadratic function of Bayesian uncertainty estimations. Note also that for larger uncertainties the epistemic component is giving consistently larger misclassification cases. This is interesting as in principle the considered problem does not admit aleatoric uncertainty (the classes do not overlap), yet in the numerical implementation the aleatoric uncertainty appears non-zero (and hence leads to different misclassification from total and epistemic uncertainties) as the result of finite precision in estimating it.


\section{Discussion}

Bayesian neural networks are probabilistic models capable of estimating the uncertainty of their predictions. They suffer, however, from the intractable distribution $p(D)$ necessary for the posterior calculation. The MCMC sampling method is computationally demanding, which limits its usage in modern business applications. Therefore, more efficient approximation techniques such as variational inference \cite{vi, seb}, Laplace approximation \cite{laplace}, and Monte Carlo Dropout \cite{dropout} are intensively developed. This problem is less severe for fundamental physics applications where the speed can be traded for the quality with a theoretical guarantee (recall that MCMC is guaranteed to converge to the proper posterior).

We also note that, in general, despite the Universal Approximation Theorem, neural network models are not perfect solutions to every predictive problem. In particular, for simpler tasks, other models, such as support vector machines or tree-based methods, can be used with better results~\cite{rfvsnn}. The goal of this paper is not to argue that neural networks are the best solution to quantum problems but rather to stress the importance and benefits brought forward by proper uncertainty estimation of the neural network model. 

In summary, we have demonstrated that Bayesian neural networks operate better than standard neural network models when training data does not cover the entire input space. In contrast to the standard methods, which tend to be overconfident, BNNs yield more conservative predictions on the parts of the input space absent in the training dataset, giving rise to meaningful uncertainty of the particular prediction.
This method was then applied to the problem of recognizing quantum contextuality. Both models were shown to learn in a similar way, but their uncertainty estimations were markedly different. Standard neural networks are confident even when they are incorrect. Uncertainty of Bayesian networks behaves in a desirable way: confident predictions tend to be correct, and uncertain ones tend to be random.
In this way uncertainty of Bayesian predictions becomes valuable figure of merit for recognising contextuality.


\subsection{Implementation details}
Experiments were implemented in python3.9. PyTorch library was used as the framework to standard train models, and Cobb's implementation of HMC algorithm NUTS \cite{cobb} was used to sample from the posterior. The code to reproduce experiments is available on GitHub.

\subsection{Acknowledgments}

        KH acknowledges the Fulbright Program and Mark Wilde for hospitality during the Fulbright scholarship at the School of Electric and Computer Engineering of the Cornell University.
    	We acknowledge partial support by the Foundation for Polish Science (IRAP project, ICTQT, contract no.\ MAB/2018/5, co-financed by EU within Smart Growth Operational Programme). The 'International Centre for Theory of Quantum Technologies' project (contract no.\ MAB/2018/5) is carried out within the International Research Agendas Programme of the Foundation for Polish Science co-financed by the European Union from the funds of the Smart Growth Operational Programme, axis IV: Increasing the research potential (Measure 4.3). 
	
	%
		
	
	


\bibliographystyle{plain.bst}
\bibliography{refs}

	




\end{document}